\documentclass[twocolumn,
               prd,nofootinbib,
               showpacs,preprintnumbers,
               amsmath,amssymb]
              {revtex4}
\usepackage{graphicx}

\begin{document}

\title{Space-time in light of K\'arolyh\'azy uncertainty relation }

\author{Michael~Maziashvili}
\email{maziashvili@hepi.edu.ge}\affiliation{Institute of High Energy Physics, 9 University St., Tbilisi 0186, Georgia\\
Andronikashvili Institute of Physics, 6 Tamarashvili St., Tbilisi 0177, Georgia }

\begin{abstract}

General relativity and quantum mechanics provide a natural explanation for the existence of dark energy with its observed value and predict its dynamics. Dark energy proves to be necessary for the existence of space-time itself and determines the rate of its stability.

\end{abstract}






\maketitle

 Due to philosophy of quantum mechanics every concept has a meaning only in terms of the experiments used to measure it, we must agree that things that cannot be measured really have no meaning in physics. So that all concepts in science require a definition based on actual, or possible (\emph{Gedanken}), experimental observations. Needless to mention that space and time are primary concepts. But, frequently we talk about space-time without asking critically how one actually measures space-time intervals. The logical foundation of the background space-time in general relativity is, in principle, the hypothesis that its characteristic laws can be verified with unlimited accuracy.  The minimal setup for observing the space-time is a clock. From the very outset it is clear that the introduction of the measuring device will alter the background space-time we are interested to measure because of its mass. Another source of disturbance is due to quantum fluctuations of the measuring device. It is easy to see that these two kinds of disturbances have the competing character as the introduction of very massive measuring device reduces its quantum jiggling but creates strong enough gravitational field disturbing the background space-time and conversely the light body is characterized with high energy fluctuations disturbing thereby the space-time. This qualitative consideration already indicates an expected imprecision in space-time structure. Endeavors to estimate the restrictions on general relativity due to uncertainty relations date back at least to the 1930 \cite{Band}. To quantify the problem let us consider a \emph{Gedankenexperiment} for time measurement. We set $\hbar=c=1$ in what follows. Further, for definiteness let our clock be a light-clock consisting of spherical mirror inside which light is bouncing, having the mass $m$ and the radius $r_c$. First let us restrict ourselves to the Minkowskian space \[ds^2=dt^2-dr^2-r^2d\Omega^2~.\] After introducing the clock the metric takes the form 

\[ds^2=\left(1-{2l_p^2m\over r}\right)dt^2- \left(1-{2l_p^2m\over r}\right)^{-1}dr^2-r^2d\Omega^2~.\] The time measured by this clock is related to the Minkowskian time as \cite{FockLL} \[t'=\left(1-{2l_p^2m\over r_c}\right)^{1/2}t~.\] From this expression one sees that the disturbance of the background metric to be small, the size of the clock should be much greater than its gravitational radius $r_c \gg 2l_p^2m$. Under this assumption for gravitational disturbance in time measurement one finds \[t'=\left(1-{l_p^2m\over r_c}\right)t~.\] The position-momentum uncertainty relation tells us that the clock is characterized by the momentum uncertainty \[\delta p={1\over 4r_c}~,\] and correspondingly its energy takes the form 

\begin{equation}\label{qcmass} Clock ~ mass \approx  m+{\delta p^2\over 2m}~.\end{equation} As the size of the clock determining its resolution time represents in itself the error during the time measurement, one finds the total imprecision in Minkowskian time measurement to have the following form 
\[ \delta t = 2r_c+{tt_p^2\over r_c}\left(m+{1\over 32 mr_c^2}\right)~.\] Minimizing this expression with respect to $m$ and $r_c$ one finds \[\delta t = \left(\sqrt{2}+{1\over\sqrt{2}}\right)\,t_p^{2/3}t^{1/3}~.\] Further refinement of the numerical factor can be done by noticing that as we are using light-clock its energy can not be less than $\pi/r_c$ giving therefore \[\delta t = 2r_c+\pi{tt_p^2\over r_c^2}~,\] which after minimization with respect to $r_c$ reads

\begin{equation}\label{timeunce}\delta t = 3\pi^{1/3}t^{1/3}t_p^{2/3}~. \end{equation} From this result immediately follows that the imprecision in Minkowskian distance measurement based on the radar method\footnote{The radar method of distance measurement is proposed in almost every textbook of general relativity, see for instance \cite{FockLL}. That is, we have a clock and by sending the light signal to some point we measure a distance by the time it takes the signal to come back to the clock.} can not be less than (apart from the numerical factor) 

\begin{equation}\label{lenun}\delta l \approx l_p^{2/3}l^{1/3}~.\end{equation} This is the result obtained by K\'arolyh\'azy long ago \cite{Karol}, which sank into oblivion until the research towards the measurement of space-time was intensified \cite{AACNvD}. In the similar way one can show that there should be a minimal resolution scale for Minkowskian space-time. Namely, to perform a measurement the radius of the clock should be greater than its gravitational radius $r_c\gtrsim 2l_p^2m$ (otherwise it will become a black hole) and since the mass of our light-clock satisfies $m\gtrsim \pi/r_c$ one gets \[r_c\gtrsim \sqrt{2\pi}\,l_p~.\] The existence of a fundamental limit to space-time resolution of the order of the Planck length was carefully analyzed in \cite{Mead}, for a review see \cite{Garay}. By differentiating the expression
\[\int\limits_{t_p}\limits^t \sqrt{1+\delta g_{00}(\xi)}\, d\xi = t-t_p+\delta t~, \] with respect to $t$, where $\delta t$ is given by Eq.(\ref{timeunce}), one finds \begin{equation}\label{flucrate}\sqrt{ 1+\delta g_{00}(t) } = 1+\pi^{1/3}\left({t_p\over t}\right)^{2/3}~. \end{equation} The above \emph{Gedankenexperiment} concerns equally well to any point of Minkowski space. The uncertainty in the length measurement via the radar method given by Eq.(\ref{lenun}), which was considered in many papers (see for instance \cite{Karol} and the paper of Ng and van Dam \cite{AACNvD}), is simply traced to the time uncertainty. In the framework of the above \emph{Gedankenexperiment} there is no motivation and it would be superfluous to assume spatial inhomogeneity and anisotropy of the metric uncertainty (\ref{flucrate}) (at least for distances $\gg l_p^2$). Due to standpoint adopted in \cite{SW, Karol,AACNvD, Mead, Garay} the unavoidable uncertainty in space-time measurement reveals its intrinsic feature to undergo fluctuations given by E.(\ref{flucrate}). Following this point of view, one expects after its emergence the Minkowski space to have the following form because of fluctuations \begin{equation}\label{quantcorrsp} ds^2= \left[1\pm\pi^{1/3}\left({t_p\over t}\right)^{2/3} \right]^2(dt^2-d\vec{x}\,^2)~, \end{equation} where $\pm$ stands for a random choice of sign with equal probability. From Eq.(\ref{quantcorrsp}) one immediately finds the energy density associated to the fluctuations of the Minkowski space   

\begin{equation}\label{classenden} \rho = {1\over 6\pi^{1/3} \left[1\pm(t_p/t)^{2/3}\right]^4\,t_p^{2/3}\, t^{10/3} }~.  \end{equation}
As we can not know the initial time to a better accuracy than $t_p$ one gets that the decaying cosmological constant

\begin{equation}\label{dcc}\Lambda \sim {l_p^{4/3}\over t^{10/3}}~,\end{equation}
 produced due to fluctuations of the Minkowskian background metric has the initial value $\sim l_p^{-2}$. Taking $t\sim H_0^{-1}$ in Eq.(\ref{dcc}), where $H_0\sim 10^{-60}l_p^{-1}$ is the present value of Hubble constant, one gets pretty small value for the cosmological constant \[\Lambda\left(t\sim H_0^{-1}\right)\sim 10^{-200}l_p^{-2}~.\] Note that the Eq.(\ref{classenden}) describes classical energy density associated to the fluctuation field, i.e., \[\rho \sim {1 \over l_p^2 }\left({d\delta g_{00}(t)\over dt }\right)^2 \sim {1 \over t_p^{2/3}t^{10/3} }~.\] Or otherwise one can derive this energy density in a more simple and clear way. From the above discussion we know that the existence of time $t$ fluctuating with the amplitude $\delta t \sim t_p^{2/3}t^{1/3}$ implies the existence of energy $E\sim \delta t^{-1}$ (the mass of the clock having the precision $\delta t$) \cite{remark}, which distributed uniformly over the volume $t^3$ gives the above expression, but this point now manifests the nature of this energy.

For estimating the proper energy budget of the space-time let us take due account of time-energy uncertainty relation. According to the K\'arolyh\'azy uncertainty relation, Eq.(\ref{timeunce}), the time $t$ can not exist to a better accuracy than $\delta t\sim t_p^{2/3}t^{1/3}$. For the time being let us forget
this relation and assume that for a given space-time there is some precision $\delta t$ and maximal time scale $t$ (the age of universe for instance) measured by this $\delta t$. As in our understanding the concept of time is intimately related to the length and there is maximal speed of motion $c=1$, the existence of $\delta t$ and $t$ immediately imply the existence of length precision $\delta l=\delta t$ and the length $l=t$. After this time-energy uncertainty relation tells us that as the age of $\delta l^3$ is $t$ its existence can not be justified with energy less than $E_{\delta l^3} \sim t^{-1}$. Correspondingly we arrive at the conclusion that: \textit{ For existence of time $t$ with the accuracy $\delta t$ there should be a homogeneously distributed energy over the region $t^3$ in space with the density}

\begin{equation}\label{darken}\rho \sim {1\over t\,\delta t^3}~.\end{equation} As the laws of nature allow us to know the time with the accuracy determined by Eq.(\ref{timeunce}) its background energy density should be

\begin{equation}\label{darkenuni}\rho \sim {1\over t_p^2\,t^2}~,\end{equation} which for $t\sim H_0^{-1}$ gives the observed value $\rho_0 \sim H_0^2/l_p^2$ \cite{PerlRiess}. And if we believe the universe was using the same laws in the past as well, then Eq.(\ref{darkenuni}) can be considered as a dynamical one for the background energy density of the universe. The time will lose its physical meaning when $\delta t  \gtrsim t$ which is tantamount to the decreasing of background energy density, Eq.(\ref{darken}), below the $\lesssim t^{-4}$. General relativity tells us that as this is the energy for existing of space-time itself its interaction with the rest of matter should be only gravitational.

Let us briefly overview the measurement of de Sitter space-time as well. It is convenient to use static coordinates for de Sitter space \cite{HE}

\[ds^2 = \left(1-{\Lambda\over 3}r^2\right)dt^2-\left(1-{\Lambda\over 3}r^2\right)^{-1}dr^2-r^2d\Omega^2~.\] At $r=0$ coordinate time coincides with the proper de Sitter time. Now let us put our light-clock at this point. In presence of clock the metric takes the form

\begin{eqnarray} ds^2  & = & \left(1-{2l_p^2m\over r}-{\Lambda\over 3}r^2\right)dt^2   \nonumber
\\ & - & \left(1-{2l_p^2m\over r}-{\Lambda\over 3}r^2\right)^{-1}dr^2-r^2d\Omega^2  ~.\end{eqnarray} Under the condition $0<3l_p^2m\sqrt{\Lambda}<1$, this metric admits event and cosmological horizons that can be written as

\begin{eqnarray}\label{radsds}
 r_g=-{2\over \sqrt{\Lambda}}\cos\left({\varphi+\pi\over 3}\right)~,&~~~~& r_{cosmo}={2\over\sqrt{\Lambda}}\cos{\varphi\over 3}~,  \nonumber\\
\\  \cos\varphi=-3l_p^2m\sqrt{\Lambda}~,&~~~~&  {\pi\over 2}< \varphi < \pi ~.\nonumber\end{eqnarray}

So, one of the conditions for small disturbance of the background metric due to clock takes the form $m\ll 1/l_p^2\sqrt{\Lambda}$. Under this condition the gravitational and cosmological horizons take approximately the standard values. The second condition we impose is the radius of the clock to be much greater than its gravitational radius and much less than the cosmological horizon $2l_p^2m\ll r_c\ll 3/\sqrt{\Lambda}$. So that, the second condition implies the first one as well. The gravitational disturbance in de Sitter time reading takes the form      

\[t'= \left(1-{l_p^2m\over r_c}-{\Lambda\over 6}r_c^2\right)t~.\] Following the above discussion, one finds that the total imprecision in de Sitter time reading is given by \[\delta t =2r_c+t \left({\pi l_p^2\over r_c^2}+{\Lambda r_c^2 \over 6}\right)~, \] which (to be minimized with respect to $r_c$) results in the equation \begin{equation}\label{dsoptsizecl}{t\Lambda \over 3}r_c^4+2r_c^3-2\pi l_p^2 t=0~,\end{equation} determining optimal value of $r_c$. Roughly, when $t\lesssim 1/\Lambda r_c $ the first term in Eq.(\ref{dsoptsizecl}) is less than or comparable to the second term and one finds the approximate solution

\begin{equation}\label{smscrel}\delta t\sim t_p^{2/3}t^{1/3}~,\end{equation} which thus holds for $t\lesssim 1/\Lambda^{3/4}t_p^{1/2}$. For $t\gg 1/\Lambda^{3/4}t_p^{1/2}$ the time uncertainty takes the form

\begin{equation}\label{largscrel}\delta t\sim t_p^{1/2}\Lambda^{-1/4}~.\end{equation} From Eqs.(\ref{smscrel},~\ref{largscrel}) one readily gets the minimal uncertainty in length measurement by replacing $tp$ and $t$ with $lp$ and $l$ respectively. By noticing that the gravitational radius of the clock given by Eq.(\ref{radsds}) is greater than that one in the Minkowskian background, $2l_p^2m$, one can repeat the above discussion to demonstrate that Planck length (time) sets the minimal resolution scale for de Sitter space as well. Furthermore, as for the performing of measurement $r_c$ should be less than the cosmological horizon, which in its turn $\lesssim 3/\sqrt{\Lambda}$ and the mass of the light-clock satisfies $m\gtrsim \pi/r_c$, with the use of relation $3l_p^2m\sqrt{\Lambda} < 1$ one finds \[3\pi l_p^2\sqrt{\Lambda} < r_c < {3\over \sqrt{\Lambda}}~.\] So it does not make any sense to talk about the de Sitter space with $\Lambda > 1/\pi l_p^2$. Hence, if the initial value of the (effective) cosmological constant is $\sim l_p^{-2}$, one can say that universe starts from the quantum foam where the fluctuations of the geometry are comparable to the geometry itself and therefore the concepts of space and time lose their meaning (the hypothesis that the universe has been created from a vacuum fluctuation was proposed in \cite{TF}). The validity condition of Eq.(\ref{smscrel}), $t\lesssim 1/H^{3/2}t_p^{1/2}$, will be satisfied during the inflation if $H\lesssim m_p/75^2 \approx 10^{-4} m_p$ (we have taken the timescale for the end of inflation $\sim 75 H^{-1}$ \cite{Mukhanov}). So that the Eq.(\ref{smscrel}) is valid during the inflation and therefore results in induced cosmological constant considered above. But as the induced cosmological constant decays very fast during the inflation it does not affect appreciably the inflationary picture.     

\begin{center}
{\bf Concluding remarks}
\end{center}

Let us start with some minor comments. The Eq.(\ref{qcmass}) is obtained under assumption $\delta v=\delta p/m \ll 1$, which enables one to use the standard position momentum uncertainty relation for the clock. But the value of mass minimizing Eq.(\ref{qcmass}), $m=\delta p/\sqrt{2}$, does not satisfy this condition. Therefore one has to take larger value of mass. Having our light-clock we surmounted this ''problem'' by using the relation $clock ~ mass \gtrsim \pi/r_c$ instead of position momentum uncertainty relation. The relation $clock ~ mass \gtrsim \pi/r_c$ simply comes from the fact that the photon confined inside the clock should have the wavelength $\lambda \lesssim 2r_c$. This relation can be also read as a statement that the clock cannot be localized more closely than its Compton wavelength.

  We followed the standpoint adopted in \cite{SW, Karol,AACNvD, Mead, Garay}, expounding that if there is an inevitable imprecision in space-time measurement it can naturally be understood as fluctuations of the space-time. This reasoning is in the spirit of quantum mechanics, that is to regard reality as that which can be observed. But attention should be payed to a rather remarkable fact that the space-time uncertainties, Eqs.(\ref{timeunce},~\ref{lenun}), depend on the length (time) scale we are measuring and correspondingly the fluctuations found in this way can be naturally ascribed to the space-time itself only in the case when it is characterized with some length (time) scale. For instance, such a natural characteristic scale associated to the space-time can be its age. In principle, what has been discussed here should also apply to horizons in different settings by rewriting the Eq.(\ref{darkenuni}) as \[\rho\sim { 1 \over l_p^2\, (Horizon ~ length)^2}~.\] Using this expression to the black holes, one finds correct value for the total energy associated to the black-hole space-time \[Black ~ hole ~ mass \sim \rho \times r_g^3 \sim {1 \over l_p^2 \, r_g^2} r_g^3 = {r_g\over l_p^2}~. \]

From Eq.(\ref{darken}) one sees that dark energy with the density $1/t\,\delta t^3$ stabilizes space-time with accuracy $\delta t$ over the length scale $t$. So if we had the present dark energy density of the order $\sim H_0^4$ it would be disastrous as it implies the fluctuations of geometry comparable to the geometry itself. The resolution of space-time described by the Eq.(\ref{darkenuni}) can not be increased as it corresponds to the minimal uncertainty given by K\'arolyh\'azy relation and the further increase of this energy density will result in a gravitational collapse of the energy associated to $l^3$ with subsequent formation of a black hole.

 To summarize the basic points of our discussion, we started with the K\'arolyh\'azy uncertainty relation allowing to estimate the fluctuations of the background metric. Estimating the energy density of fluctuations corresponding to a given time scale $t$ \[ \rho \sim { 1 \over t_p^{2/3}t^{10/3} }~,\] one finds that the minimal cell of space over the length scale $t$ allowed by the K\'arolyh\'azy uncertainty relation, $\delta t^3 \sim t_p^2\,t$, carries energy \[E(\delta t^3)_{classical}\sim {t_p^{4/3}\over t^{7/3}}~.\] Further insight comes from the time-energy uncertainty relation. Namely, time-energy uncertainty relation tells us that as the age of the cell $\delta t ^3$ is $t$ its energy can not be smaller than \[E(\delta t^3)_{quantum}\sim {1\over t}~.\] The quantum corrected expression for the energy density of fluctuations \[\rho \sim {1 \over t_p^2\,t^2}~,\] which for $t \sim H_0^{-1}$ is in good agreement with the observations $\rho_0 \sim H_0^2/l_p^2$, determines the whole dynamics of the dark energy density during the evolution of the universe. The conceptual point coming from this discussion is that:\textit{ For its existence space-time needs some energy which according to the general relativity should manifest itself through the gravitational interaction only and it is dark energy we observe today.}

\begin{acknowledgments}

The author wishes to acknowledge the hospitality of the \emph{Department of Astroparticle Physics and Cosmology, Arnold Sommerfeld Center for Theoretical Physics (ASC), Ludwig Maximilians Universit\"at M\"unchen} where this work was done and thanks Professor V.~Mukhanov for invitation to \emph{ASC}, V.~Vanchurin and S.~Winitzki for valuable comments and discussions. The work was supported by the \emph{DAAD Fellowship}, \emph{INTAS Fellowship for Young Scientists} and the \emph{Georgian President Fellowship for Young Scientists}.

\end{acknowledgments}


\begin{thebibliography}{10}
\bibitem{remark}

This estimate can be done simply by the time-energy uncertainty relation as well.
\end{thebibliography}
\end{document}